\documentclass[sigconf]{acmart}

\AtBeginDocument{%
  \providecommand\BibTeX{{%
    \normalfont B\kern-0.5em{\scshape i\kern-0.25em b}\kern-0.8em\TeX}}}

\setcopyright{acmcopyright}
\copyrightyear{2023}
\acmYear{2023}
\acmDOI{XXXXXXX.XXXXXXX}

\acmBooktitle{Proceedings of ArXiv}
\usepackage{xspace}

\newcommand{\README}{\texttt{README}\xspace}
\newcommand{\autofl}[0]{\texttt{AutoFL}\xspace}
\newcommand{\lbl}[1]{`\textit{#1}'\xspace}
\newcommand*{\ie}{i.e.,\@\xspace}
\newcommand*{\eg}{e.g.,\@\xspace}

\usepackage{listings}
\begin{document}

\title{AutoFL: A Tool for Automatic Multi-granular Labelling of Software Repositories}

\author{Cezar Sas}
\email{c.a.sas@rug.nl}
\orcid{0000-0002-3018-0140}
\affiliation{%
  \institution{University of Groningen}
  \streetaddress{Nijenborgh 9}
  \city{Groningen}
  \country{Netherlands}
  \postcode{9747 AG}
}
\author{Andrea Capiluppi}
\email{a.capiluppi@rug.nl}
\orcid{0000-0001-9469-6050}
\affiliation{%
  \institution{University of Groningen}
  \streetaddress{Nijenborgh 9}
  \city{Groningen}
  \country{Netherlands}
  \postcode{9747 AG}
}


\begin{abstract}

Software comprehension, especially of new code bases, is time consuming for developers, especially in large projects with multiple functionalities spanning various domains.
One strategy to reduce this effort involves annotating files with meaningful labels that describe the functionalities contained. However, prior research has so far focused on classifying the whole project using \README files as a proxy, resulting in little information gained for the developers.

Our objective is to streamline the labelling of files with the correct application domains using source code as input. To achieve this, in prior work, we evaluated the ability to annotate files automatically using a weak labelling approach. 

This paper presents \autofl, a tool for automatically labelling software repositories from source code. \autofl allows multi-granular annotations including: \textit{file}, \textit{package}, and \textit{project} -level. 

We provide an overview of the tool's internals, present an example analysis for which \autofl can be used, and discuss limitations and future work.

\end{abstract}





\received{29 January 2024}
\received[revised]{-}
\received[accepted]{-}

\maketitle

\section{Introduction}

The increased amount of code created during software development has led to the expansion and more considerable complexity of codebases. It is estimated that developers devote approximately 70\% of their time to software comprehension~\cite{minelli2015time, xia2017measuring}. Streamlining this process holds the potential to accelerate the integration of new developers into a project and reduce the time spent on software maintenance, which also demands substantial investment in program comprehension~\cite{xia2017measuring}. As a result, there is a growing need for tools that can enhance software comprehension, facilitating a faster and more complete understanding of a project's semantic content. 

Prior work has focused its efforts on augmenting comprehension by annotating software projects with project-level labels, which are indicative of the project's application domains~\cite{izadi2020topic,zhang2019HiGitClass,rocco2023hybridrec,sipio2020naive}.
In~\cite{rocco2023hybridrec, sipio2020naive}, the authors used Na\"ive Bayesian Networks, and the \README file accompanying a project, to recommend labels for GitHub repositories. More neural network-based approaches were attempted: for instance, the work presented in~\cite{izadi2020topic} employs BERT, while the approaches used in~\cite{leclair2018neural, ohashi2019cnn_code} use a convolutional neural network and long short-term memory for the classification, respectively.

Most of these works examine \README files, which, while offering a general insight into the functionalities of a software project, delving into the source code provides a more precise understanding. Existing methods, like those presented in~\cite{kuhn2007semantic}, concentrate on topic modeling. Despite this, the extracted topics often remain ambiguous, necessitating comprehension from developers.

To address the mentioned issue, one can implement file-level classification, (\ie annotate the functionality of each file); this ensures the consideration of all distinct functionalities within the source code, not solely those, if any, highlighted in the \README. However, we note a scarcity of research in this type of classification, mainly due to the lack of data necessary to train models. However, in recent years, machine learning research has addressed this issue with weak-supervision~\cite{zhang2022weak}. This approach uses heuristics, distant supervision, or domain expert knowledge to automatically create weakly labelled (\ie noisy, not gold standard) training data for machine learning (ML) models. The annotation methods are implemented programmatically using so-called labelling functions (LFs). 

Building on this concept, our prior work~\cite{sas2023multigranular} investigated file-level annotation using weak-labelling. We proposed a multi-granular approach for the annotation at file, package, and project-level. Based on this work, we are now developing \autofl~\cite{Sas_AutoFL_2023}, a tool for multi-granular software repositories annotation. 



\begin{figure*}[htbp!]
    \centering
    \includegraphics[width=\textwidth]{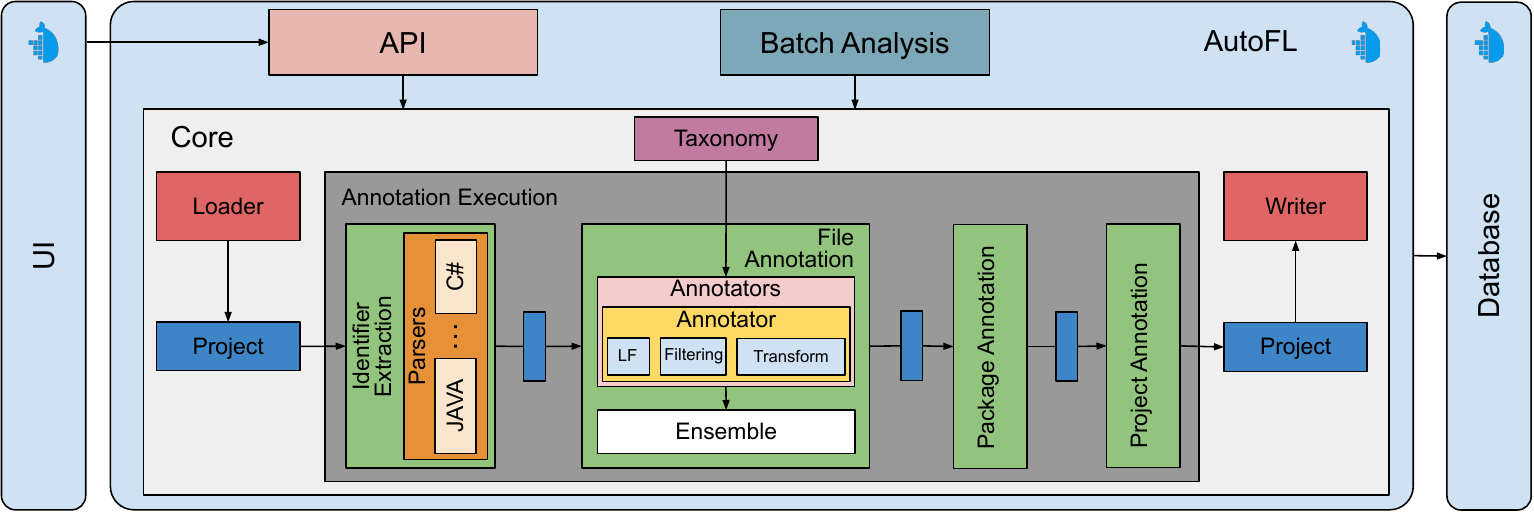}
    \caption{Abstract architecture of the \autofl. There are three modules, each running in a Docker container. The annotation is done in the core component. First, the content is parsed to extract the identifiers; then, the files are labelled using different annotators. Optionally, the package and project-level annotations are executed.}
    \label{fig:architecture}
\end{figure*}

\section{\texttt{A\MakeLowercase{uto}FL}}
\label{sec:_autofl}
\subsection{Architecture}

\autofl's architecture comprises three parts, as shown in Figure~\ref{fig:architecture}. The three components are running as separate Docker containers:

\begin{enumerate}
    \item \textbf{UI} module: offers the dashboard to visualize per-project analysis results. 
    \item \textbf{\autofl} module is the main module; it performs the annotations and offers both API and batch analysis. 
    \item \textbf{Database} module: stores the results for the analysis.
\end{enumerate}

\subsubsection{UI Module}
The UI module allows an easy access to the result for the annotations. It comprises different views for the annotations at the various levels. The UI is developed using Streamlit\footnote{\href{https://streamlit.io/}{https://streamlit.io/}}, a Python framework for the creation of data-oriented web apps. This makes it easy to create personalized interfaces and expansions. Connects to the \autofl module via the provided APIs.

\subsubsection{\autofl Module}

It is responsible for the annotation and can be accessed using both API and batch analysis. The \textbf{Core} component comprises several key sub-modules, of which interaction can be seen in Figure~\ref{fig:architecture}.

All sub-components can be can be customized via YAML files using Hydra\footnote{\href{https://hydra.cc/}{https://hydra.cc/}} configuration manager. The configuration allows customized execution logic by instantiating classes defined in the YAML files. Examples of what can be customized include the classes used for the \textit{Annotators}, \textit{Ensemble}, and various other parameters. We present the functionalities offered by each component in the order of execution.

\textbf{Taxonomy}: stores information about the taxonomy used for the labelling. This includes the labels list with the associated label ID and, optionally, keywords associated with each label. The keywords have weights, which are necessary when using keyword-based LFs. 

\textbf{Loader}: creates the \textbf{Project} class used through the annotation pipeline. It is a custom class and depends on the dataset format.

\textbf{Project}: contains all the project-related details, including the project's name, the remote URL of the repository hosting the source code, and a list of versions based on the commit id. Each version has a list of files.

\textbf{Annotation Execution}: prepares the project for analysis by pulling or checking out the correct version from the version control system. It executes 4 steps (listed below) required for the multi-granular annotation. 

1 -- \textbf{Identifier Extraction}: it parses the file's content to extract the identifiers. It utilizes language-specific \textbf{Parsers} developed with \texttt{tree-sitter}. Presently, parsers are available for languages such as Java, Python, C, C++, and C\#. 

2 -- \textbf{File Annotation}: manages the annotation of files by running all the annotators and the ensemble. Each \textbf{Annotator} in the \textbf{Annotators} subcomponent uses the file's content and the taxonomy to annotate the file. The annotation is achieved as follows:
\begin{itemize}
 \item \textbf{LF}: defines the logic used to perform the annotation of the files. Currently, \autofl supports both keyword-based and similarity-based LFs, but it is possible to define custom ones. The output is a vector representing the probability distribution over a set of $m$ variables (\ie the number of labels in the taxonomy). For example, the keyword-based LFs, will check for each label how many keywords belong to that label are present in the file, then TFIDF weighting is applied;
 \item  \textbf{Filtering}: checks if the annotation meets specific criteria. \autofl uses the Jensen–Shannon divergence~\cite{endres2003metric} to identify noisy annotations; however, custom functions can also be implemented. The file is marked as `\textit{unannotated}' if the criteria are unmet;
 \item  \textbf{Transformation}: refines the output of LF by selecting the best label or those surpassing a custom probability threshold. Examples include the most likely label, or top-n labels. As for the other parts, custom versions can be implemented.
\end{itemize}
The annotators' result is then used by the \textbf{Ensemble} subcomponent. It combines all the annotations into a single probability vector using approaches like voting or averaging.

3 -- \textbf{Package Annotation}\footnote{Steps 3, and 4 are optional, with the package annotation only possible for languages with package-like concepts (\eg modules in Python)}: annotates the packages in a project. It does so by averaging annotations from the package constituent files that are annotated (\ie passed the filtering check). 

4 -- \textbf{Project Annotation}: annotates a project by averaging annotations from all annotated files within the project.

\textbf{Writer}: stores the analysis results in a database (\eg PostgreSQL) or a file (\eg JSON). A custom class can be implemented for other DBs or file formats.

\subsubsection{Database}
For the database (DB), we use PostgreSQL\footnote{\href{https://www.postgresql.org/}{https://www.postgresql.org/}}. Projects are stored in a single table, and the table schema is defined as follows (in \textbf{boldface} are primary key columns):

\begin{itemize}
    \item \textbf{name}: the name of the project;
    \item \textbf{version\_sha}: commit id in the repository;
    \item version\_num: version position in the commit history;
    \item \textbf{config}: Hydra configuration for the specific analysis;
    \item project: a JSON object containing the project's information;
    \item version: a JSON object containing the files and their annotations for the specific version.
\end{itemize}

While a NoSQL DB like MongoDB would have been better suited given our `\textit{no-schema}'  design, the size of each document can easily be above the hard limit of 16MB in MongoDB. In contrast, PostgreSQL's JSON column has a limit of 256MB while still allowing for queries over the document.

\subsection{Current Use Cases}
In our prior work~\cite{sas2023multigranular}, we used the \textbf{Core} part of the tool to create an annotated dataset of around 2,600 Java projects in 267 application domains. This dataset is publicly available\footnote{\href{https://zenodo.org/record/7943882}{https://zenodo.org/record/7943882}} and, among others, can be used as a source of weak supervision for training ML models.

The qualitative evaluation of our approach can be found in detail in the original work: the main results are in Table~\ref{tab:eval}. The evaluation was performed with automated (A) metrics and human evaluation (H), as the ground truth is only available at the project level.

We can see that the recall at the project-level is relatively high; however, the ability to discover, on average, 3 \textit{new labels} per project is a better example of the need for file-level annotations. The performance is still high at the package and file level, considering the large number of labels (267). It is important to note that the results are not for a classification task but to evaluate the quality of the LFs to annotate the dataset. We refer readers to our previous work~\cite{sas2023multigranular} for details regarding the configuration used to get these results.

\begin{table}[htbp!]
    \caption{Summary of \autofl performances from~\cite{sas2023multigranular}.}
    \centering
    \begin{tabular}{llcc}
    \toprule
    \textbf{Level} & \textbf{Metric} & \textbf{Score} & \textbf{Type} \\
    \midrule
     Project & Recall@10     &  70\% & A \\
     Project & \# New Found Labels    &  3.24 & H \\
     Package & SuccessRate@3 &  57\% & H \\
     File & SuccessRate@3    &  50\% & H\\
     \bottomrule
    \end{tabular}
    \label{tab:eval}
\end{table}

At the moment, the tool userbase consists of a couple of PhD Students who are using the tool in their research in software engineering to gain more insights from the metrics with respect to the application domain. 

\subsection{Future Use Cases}
\autofl can be helpful in future use cases. Firstly, it can be used to perform general large-scale annotations of source code. These annotations can be further used for empirical software engineering researcher to contextualize their result based on the application domain~\cite{capiluppi2019relevance}; by analysing application domains separately, instead of all together, the results can give more insights and used for better domain-driven software design~\cite{evans2004domain}. 

Furthermore, \autofl can assist in improving reusability~\cite{sandhu2021reuse, barros2019reuse}, as the annotated packages can be extracted and reused in new projects that need the included functionality. 

Finally, \autofl can assist with the comprehension task by providing multi-granular annotations; developers can better comprehend what functionality a codebase offers and gather more knowledge about the inner details of a library that they might already be familiar with. This use case will be showcased in Section~\ref{sec:analysis}.

\subsection{Limitations}
Considering that the development of AutoFL is still in early stages, the tool has some limitations. The first limitation is due to \autofl's \textit{heuristic-based} nature: annotations do not rely on a classification model but on heuristics. While this approach has shown promise, it lacks the generality of a machine learning model, due to the limited research and data in this area. Our team is actively working to address these challenges.

Another limitation, based on \textit{Java Keywords}, arises from the taxonomy's dependence on keywords exclusively extracted from Java projects. This specificity reduces the generalization ability of the annotations to other programming languages, especially when considering the diverse preferences within various communities (\eg machine learning practitioners preferring Python).

The \textit{database model} also presents a limitation. The single-table design makes it hard to perform more complex analytics and hinders the visualization of aggregated results from multiple projects. This could be beneficial for reusing software components or comparing different annotation configurations when, for example, a new LF is implemented.

Additional, albeit minor, limitation involves \textit{UI settings}, and the limited options to customize the execution from the UI. While \autofl is easily configurable with Hydra, improvements are necessary in the UI to showcase these settings effectively.

Lastly, \autofl lacks \textit{IDE integration}, which makes it more challenging for developers to use \autofl during development.

\section{Example of Analysis}
\label{sec:analysis}

\autofl can assist with comprehending what functionality a codebase offers and also gathering additional knowledge about the inner details of a library that the developer might already be familiar with. This info can also assist with reusing sub-components.

For our showcase, we picked Pumpernickel\footnote{\href{https://github.com/mickleness/pumpernickel}{https://github.com/mickleness/pumpernickel}}, a small library for UI development written in Java.
Looking at the project's GitHub pages, we see that Pumpernickel is a \lbl{UI} library: this is the only application domain label assigned by its developers. While it is sufficiently descriptive to get a general view of the functionalities offered, it does not give enough details of all parts of the software. Furthermore, the repository lacks a descriptive \README, making it impossible for previous tools to classify just at the project-level.

\begin{figure}[htbp!]
    \centering
    \includegraphics[width=.95\linewidth]{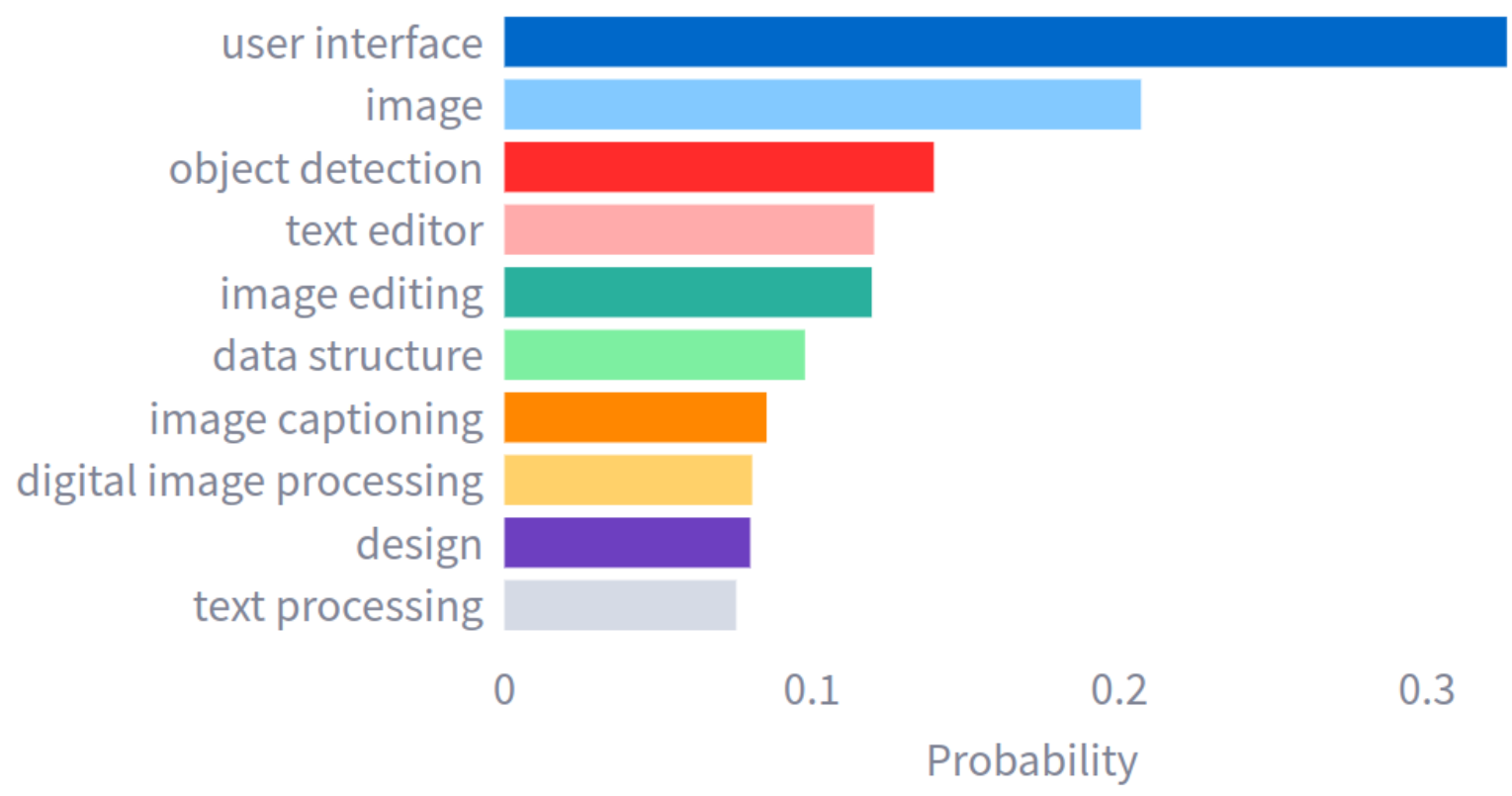}
    \caption{Project-level UI showing the top 10 annotations.}
    \label{fig:Project}
\end{figure}

By running \autofl, and checking the identified project-level labels (Figure~\ref{fig:Project}), we discover new labels that provide more information about the project. \lbl{Image}, \lbl{Text Editor}, \lbl{Digital Image Processing}, and \lbl{Design} are all relevant labels that add information on what the project offers.

Due to the heuristic-based approach in \autofl, not all the identified project-level labels are acceptable: \lbl{Object Detection} and \lbl{Image Captioning} are ML application domains, and not relevant for Pumpernickel. This mishap is due to an overlap between `image' keywords contained in both correct (\eg \lbl{Digital Image Processing}) and incorrect (\eg \lbl{Image Captioning}) labels.

\begin{figure}[htbp!]
    \centering
    \includegraphics[width=\linewidth]{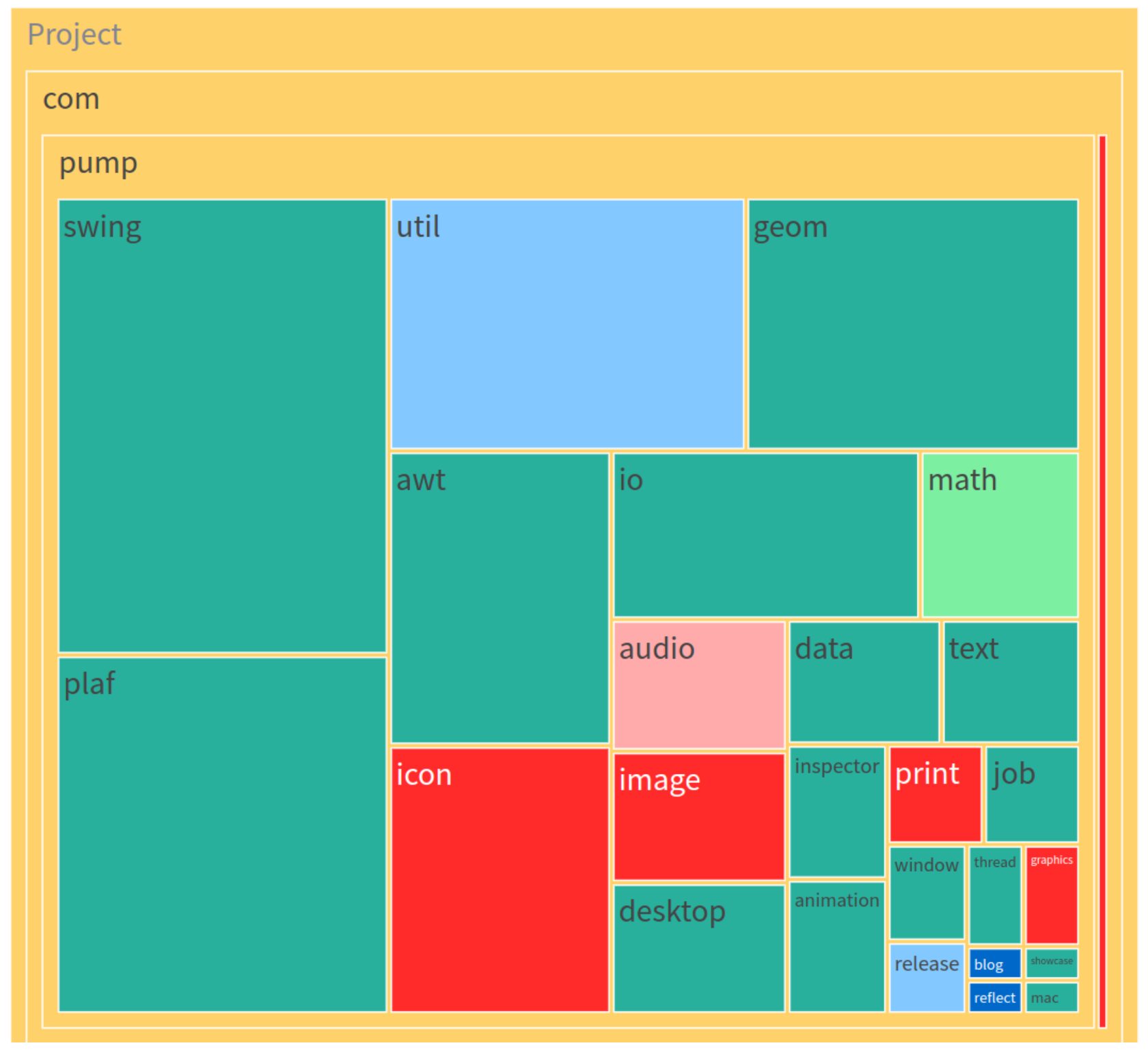}
    \caption{Package-level annotations UI. The rectangles are the packages, their size is the number of files contained, and the colour is the label. Hovering in the UI shows the label.}
    \label{fig:package}
\end{figure}

Considering this project at the package-level, \autofl helps in detecting where these functionalities are. Figure~\ref{fig:package} showcases the annotated packages for the project. As expected, the majority of packages are annotated with \lbl{User Interface} (in dark green, \eg \texttt{swing}), while others, in red (\eg \texttt{image}), with the \lbl{Image} label.

\begin{figure}[htbp!]
    \centering
    \includegraphics[width=\linewidth]{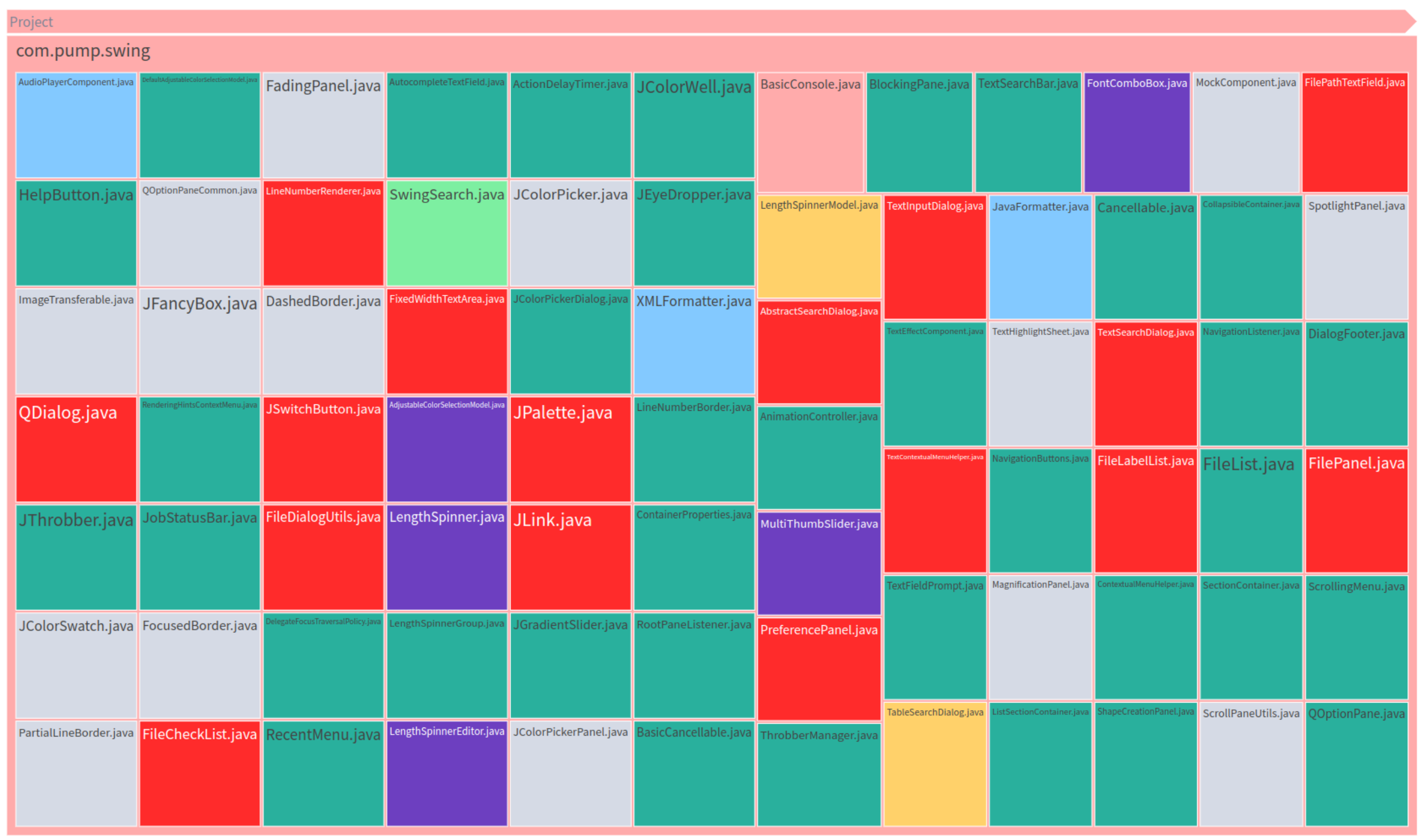}
    \caption{File-level annotations UI for the \texttt{swing} package. Rectangles are files, and colour is the label, which differs from previous plots. Hovering in the UI shows the label.} 
    \label{fig:file}
\end{figure}

The information is more detailed but noisier at the file-level (Figure~\ref{fig:file}), but it can still be used to identify files that might not fit in the intended package functionality.

\section{Future Work}
\label{sec:future}
Future work will focus on performing industry validation of the tool's ability to assist developers in better understanding codebases. Furthermore, it will also focus on addressing the current limitations.

First, we want to add a machine-learning model to address the \textit{heuristic-based} limitation. Using the tool's generated data, a weakly supervised~\cite{zhang2022weak} ML model can be trained. 

We also plan to address the \textit{Java keyword} limitation. Extracting the keywords from all the supported languages requires minor adjustments, as the pipeline used for the current keywords from Java projects can be used for all programming languages using specific parsers. 

To address the \textit{database model} limitation, data normalisation~\cite{william1983normalization} processes will be required. This will reduce the data's redundancy, allowing for easier data aggregation to perform analyses.

More work on the UI is required for the \textit{UI settings}; however, no significant changes to the core are required as Hydra configs can be easily overwritten at runtime.

Lastly, for the \textit{IDE integration}, more work needs to be placed on efficiently showcasing the information to the developers. 

\section{Conclusions}
\label{sec:_conclusion}
We presented \autofl, a tool for multi-granular labelling of software projects. The tool aims to assist with software comprehension by assigning a label at the file, package, and project-level. The tool is customizable and supports various programming languages. Furthermore, it allows for single analysis with the UI and batch analysis for empirical research. 

\section{Tool Availability}
\autofl is available for researchers and practitioners on GitHub\footnote{\href{https://github.com/SasCezar/AutoFL/}{https://github.com/SasCezar/AutoFL/}}.  Furthermore, it is also available in a persistent online repository Zenodo~\cite{Sas_AutoFL_2023}. Lastly, a video walkthrough is available\footnote{\href{https://youtu.be/ZYWZdYcip2A}{https://youtu.be/ZYWZdYcip2A}}.


\bibliographystyle{ACM-Reference-Format}
\bibliography{sample-base}


\begin{thebibliography}{18}


\ifx \showCODEN    \undefined \def \showCODEN     #1{\unskip}     \fi
\ifx \showDOI      \undefined \def \showDOI       #1{#1}\fi
\ifx \showISBNx    \undefined \def \showISBNx     #1{\unskip}     \fi
\ifx \showISBNxiii \undefined \def \showISBNxiii  #1{\unskip}     \fi
\ifx \showISSN     \undefined \def \showISSN      #1{\unskip}     \fi
\ifx \showLCCN     \undefined \def \showLCCN      #1{\unskip}     \fi
\ifx \shownote     \undefined \def \shownote      #1{#1}          \fi
\ifx \showarticletitle \undefined \def \showarticletitle #1{#1}   \fi
\ifx \showURL      \undefined \def \showURL       {\relax}        \fi
\providecommand\bibfield[2]{#2}
\providecommand\bibinfo[2]{#2}
\providecommand\natexlab[1]{#1}
\providecommand\showeprint[2][]{arXiv:#2}

\bibitem[Barros-Justo et~al\mbox{.}(2019)]%
        {barros2019reuse}
\bibfield{author}{\bibinfo{person}{José~L. Barros-Justo}, \bibinfo{person}{Fabiane~B.V. Benitti}, {and} \bibinfo{person}{Santiago Matalonga}.} \bibinfo{year}{2019}\natexlab{}.
\newblock \showarticletitle{Trends in software reuse research: A tertiary study}.
\newblock \bibinfo{journal}{\emph{Computer Standards \& Interfaces}}  \bibinfo{volume}{66} (\bibinfo{year}{2019}), \bibinfo{pages}{103352}.
\newblock
\showISSN{0920-5489}
\urldef\tempurl%
\url{https://doi.org/10.1016/j.csi.2019.04.011}
\showDOI{\tempurl}


\bibitem[Capiluppi and Ajienka(2019)]%
        {capiluppi2019relevance}
\bibfield{author}{\bibinfo{person}{Andrea Capiluppi} {and} \bibinfo{person}{Nemitari Ajienka}.} \bibinfo{year}{2019}\natexlab{}.
\newblock \showarticletitle{The relevance of application domains in empirical findings}. In \bibinfo{booktitle}{\emph{Proceedings of the 2nd International Workshop on Software Health, SoHeal@ICSE 2019, Montreal, QC, Canada, May 28, 2019}}, \bibfield{editor}{\bibinfo{person}{Bram Adams}, \bibinfo{person}{Eleni Constantinou}, \bibinfo{person}{Tom Mens}, \bibinfo{person}{Kate Stewart}, {and} \bibinfo{person}{Gregorio Robles}} (Eds.). \bibinfo{publisher}{{IEEE} / {ACM}}, \bibinfo{pages}{17--24}.
\newblock
\urldef\tempurl%
\url{https://dl.acm.org/citation.cfm?id=3355304}
\showURL{%
\tempurl}


\bibitem[Endres and Schindelin(2003)]%
        {endres2003metric}
\bibfield{author}{\bibinfo{person}{Dominik~Maria Endres} {and} \bibinfo{person}{Johannes~E. Schindelin}.} \bibinfo{year}{2003}\natexlab{}.
\newblock \showarticletitle{A new metric for probability distributions}.
\newblock \bibinfo{journal}{\emph{{IEEE} Trans. Inf. Theory}} \bibinfo{volume}{49}, \bibinfo{number}{7} (\bibinfo{year}{2003}), \bibinfo{pages}{1858--1860}.
\newblock
\urldef\tempurl%
\url{https://doi.org/10.1109/TIT.2003.813506}
\showDOI{\tempurl}


\bibitem[Evans(2004)]%
        {evans2004domain}
\bibfield{author}{\bibinfo{person}{Eric Evans}.} \bibinfo{year}{2004}\natexlab{}.
\newblock \bibinfo{booktitle}{\emph{Domain-driven design: tackling complexity in the heart of software}}.
\newblock \bibinfo{publisher}{Addison-Wesley Professional}.
\newblock


\bibitem[Izadi et~al\mbox{.}(2021)]%
        {izadi2020topic}
\bibfield{author}{\bibinfo{person}{Maliheh Izadi}, \bibinfo{person}{Abbas Heydarnoori}, {and} \bibinfo{person}{Georgios Gousios}.} \bibinfo{year}{2021}\natexlab{}.
\newblock \showarticletitle{Topic recommendation for software repositories using multi-label classification algorithms}.
\newblock \bibinfo{journal}{\emph{Empirical Software Engineering}} \bibinfo{volume}{26}, \bibinfo{number}{5} (\bibinfo{year}{2021}), \bibinfo{pages}{93}.
\newblock
\urldef\tempurl%
\url{https://doi.org/10.1007/s10664-021-09976-2}
\showDOI{\tempurl}


\bibitem[Kent(1983)]%
        {william1983normalization}
\bibfield{author}{\bibinfo{person}{William Kent}.} \bibinfo{year}{1983}\natexlab{}.
\newblock \showarticletitle{A Simple Guide to Five Normal Forms in Relational Database Theory}.
\newblock \bibinfo{journal}{\emph{Commun. ACM}} \bibinfo{volume}{26}, \bibinfo{number}{2} (\bibinfo{date}{feb} \bibinfo{year}{1983}), \bibinfo{pages}{120–125}.
\newblock
\showISSN{0001-0782}
\urldef\tempurl%
\url{https://doi.org/10.1145/358024.358054}
\showDOI{\tempurl}


\bibitem[Kuhn et~al\mbox{.}(2007)]%
        {kuhn2007semantic}
\bibfield{author}{\bibinfo{person}{Adrian Kuhn}, \bibinfo{person}{Stéphane Ducasse}, {and} \bibinfo{person}{Tudor Gîrba}.} \bibinfo{year}{2007}\natexlab{}.
\newblock \showarticletitle{Semantic clustering: Identifying topics in source code}.
\newblock \bibinfo{journal}{\emph{Information and Software Technology}} \bibinfo{volume}{49}, \bibinfo{number}{3} (\bibinfo{year}{2007}), \bibinfo{pages}{230--243}.
\newblock
\showISSN{0950-5849}
\urldef\tempurl%
\url{https://doi.org/10.1016/j.infsof.2006.10.017}
\showDOI{\tempurl}
\newblock
\shownote{12th Working Conference on Reverse Engineering}.


\bibitem[LeClair et~al\mbox{.}(2018)]%
        {leclair2018neural}
\bibfield{author}{\bibinfo{person}{Alexander LeClair}, \bibinfo{person}{Zachary Eberhart}, {and} \bibinfo{person}{Collin McMillan}.} \bibinfo{year}{2018}\natexlab{}.
\newblock \showarticletitle{Adapting Neural Text Classification for Improved Software Categorization}. In \bibinfo{booktitle}{\emph{2018 {IEEE} International Conference on Software Maintenance and Evolution, {ICSME} 2018, Madrid, Spain, September 23-29, 2018}}. \bibinfo{publisher}{{IEEE} Computer Society}, \bibinfo{pages}{461--472}.
\newblock
\urldef\tempurl%
\url{https://doi.org/10.1109/ICSME.2018.00056}
\showDOI{\tempurl}


\bibitem[Minelli et~al\mbox{.}(2015)]%
        {minelli2015time}
\bibfield{author}{\bibinfo{person}{Roberto Minelli}, \bibinfo{person}{Andrea Mocci}, {and} \bibinfo{person}{Michele Lanza}.} \bibinfo{year}{2015}\natexlab{}.
\newblock \showarticletitle{I know what you did last summer: an investigation of how developers spend their time}. In \bibinfo{booktitle}{\emph{Proceedings of the 2015 {IEEE} 23rd International Conference on Program Comprehension, {ICPC} 2015, Florence/Firenze, Italy, May 16-24, 2015}}, \bibfield{editor}{\bibinfo{person}{Andrea~De Lucia}, \bibinfo{person}{Christian Bird}, {and} \bibinfo{person}{Rocco Oliveto}} (Eds.). \bibinfo{publisher}{{IEEE} Computer Society}, \bibinfo{pages}{25--35}.
\newblock
\urldef\tempurl%
\url{https://doi.org/10.1109/ICPC.2015.12}
\showDOI{\tempurl}


\bibitem[Ohashi and Watanobe(2019)]%
        {ohashi2019cnn_code}
\bibfield{author}{\bibinfo{person}{Hiroki Ohashi} {and} \bibinfo{person}{Yutaka Watanobe}.} \bibinfo{year}{2019}\natexlab{}.
\newblock \showarticletitle{Convolutional Neural Network for Classification of Source Codes}. In \bibinfo{booktitle}{\emph{13th {IEEE} International Symposium on Embedded Multicore/Many-core Systems-on-Chip, MCSoC 2019, Singapore, Singapore, October 1-4, 2019}}. \bibinfo{publisher}{{IEEE}}, \bibinfo{pages}{194--200}.
\newblock
\urldef\tempurl%
\url{https://doi.org/10.1109/MCSoC.2019.00035}
\showDOI{\tempurl}


\bibitem[Rocco et~al\mbox{.}(2023)]%
        {rocco2023hybridrec}
\bibfield{author}{\bibinfo{person}{Juri~Di Rocco}, \bibinfo{person}{Davide~Di Ruscio}, \bibinfo{person}{Claudio~Di Sipio}, \bibinfo{person}{Phuong~T. Nguyen}, {and} \bibinfo{person}{Riccardo Rubei}.} \bibinfo{year}{2023}\natexlab{}.
\newblock \showarticletitle{HybridRec: {A} recommender system for tagging GitHub repositories}.
\newblock \bibinfo{journal}{\emph{Appl. Intell.}} \bibinfo{volume}{53}, \bibinfo{number}{8} (\bibinfo{year}{2023}), \bibinfo{pages}{9708--9730}.
\newblock
\urldef\tempurl%
\url{https://doi.org/10.1007/s10489-022-03864-y}
\showDOI{\tempurl}


\bibitem[Sandhu and Batth(2021)]%
        {sandhu2021reuse}
\bibfield{author}{\bibinfo{person}{Amandeep~Kaur Sandhu} {and} \bibinfo{person}{Ranbir~Singh Batth}.} \bibinfo{year}{2021}\natexlab{}.
\newblock \showarticletitle{A Hybrid approach to identify Software Reusable Components in Software Intelligence}. In \bibinfo{booktitle}{\emph{2021 2nd International Conference on Intelligent Engineering and Management (ICIEM)}}. \bibinfo{pages}{353--356}.
\newblock
\urldef\tempurl%
\url{https://doi.org/10.1109/ICIEM51511.2021.9445378}
\showDOI{\tempurl}


\bibitem[Sas and Capiluppi(2023)]%
        {Sas_AutoFL_2023}
\bibfield{author}{\bibinfo{person}{Cezar Sas} {and} \bibinfo{person}{Andrea Capiluppi}.} \bibinfo{year}{2023}\natexlab{}.
\newblock \bibinfo{booktitle}{\emph{{AutoFL}}}.
\newblock
\urldef\tempurl%
\url{https://doi.org/10.5281/zenodo.10255367}
\showDOI{\tempurl}


\bibitem[Sas and Capiluppi(2024)]%
        {sas2023multigranular}
\bibfield{author}{\bibinfo{person}{Cezar Sas} {and} \bibinfo{person}{Andrea Capiluppi}.} \bibinfo{year}{2024}\natexlab{}.
\newblock \showarticletitle{Multi-granular Software Annotation using File-level Weak Labelling}.
\newblock \bibinfo{journal}{\emph{Empirical Software Engineering}} \bibinfo{volume}{29}, \bibinfo{number}{1} (\bibinfo{year}{2024}), \bibinfo{pages}{12}.
\newblock
\urldef\tempurl%
\url{https://doi.org/10.1007/s10664-023-10423-7}
\showDOI{\tempurl}


\bibitem[Sipio et~al\mbox{.}(2020)]%
        {sipio2020naive}
\bibfield{author}{\bibinfo{person}{Claudio~Di Sipio}, \bibinfo{person}{Riccardo Rubei}, \bibinfo{person}{Davide~Di Ruscio}, {and} \bibinfo{person}{Phuong~Thanh Nguyen}.} \bibinfo{year}{2020}\natexlab{}.
\newblock \showarticletitle{A Multinomial Na{\"{\i}}ve Bayesian {(MNB)} Network to Automatically Recommend Topics for GitHub Repositories}.
\newblock In \bibinfo{booktitle}{\emph{{EASE} '20: Evaluation and Assessment in Software Engineering, Trondheim, Norway, April 15-17, 2020}}, \bibfield{editor}{\bibinfo{person}{Jingyue Li}, \bibinfo{person}{Letizia Jaccheri}, \bibinfo{person}{Torgeir Dings{\o}yr}, {and} \bibinfo{person}{Ruzanna Chitchyan}} (Eds.). \bibinfo{publisher}{{ACM}}, \bibinfo{pages}{71--80}.
\newblock
\urldef\tempurl%
\url{https://doi.org/10.1145/3383219.3383227}
\showDOI{\tempurl}


\bibitem[Xia et~al\mbox{.}(2017)]%
        {xia2017measuring}
\bibfield{author}{\bibinfo{person}{Xin Xia}, \bibinfo{person}{Lingfeng Bao}, \bibinfo{person}{David Lo}, \bibinfo{person}{Zhenchang Xing}, \bibinfo{person}{Ahmed~E Hassan}, {and} \bibinfo{person}{Shanping Li}.} \bibinfo{year}{2017}\natexlab{}.
\newblock \showarticletitle{Measuring program comprehension: A large-scale field study with professionals}.
\newblock \bibinfo{journal}{\emph{IEEE Transactions on Software Engineering}} \bibinfo{volume}{44}, \bibinfo{number}{10} (\bibinfo{year}{2017}), \bibinfo{pages}{951--976}.
\newblock


\bibitem[Zhang et~al\mbox{.}(2022)]%
        {zhang2022weak}
\bibfield{author}{\bibinfo{person}{Jieyu Zhang}, \bibinfo{person}{Cheng{-}Yu Hsieh}, \bibinfo{person}{Yue Yu}, \bibinfo{person}{Chao Zhang}, {and} \bibinfo{person}{Alexander Ratner}.} \bibinfo{year}{2022}\natexlab{}.
\newblock \showarticletitle{A Survey on Programmatic Weak Supervision}.
\newblock \bibinfo{journal}{\emph{CoRR}}  \bibinfo{volume}{abs/2202.05433} (\bibinfo{year}{2022}).
\newblock
\showeprint[arXiv]{2202.05433}
\urldef\tempurl%
\url{https://arxiv.org/abs/2202.05433}
\showURL{%
\tempurl}


\bibitem[Zhang et~al\mbox{.}(2019)]%
        {zhang2019HiGitClass}
\bibfield{author}{\bibinfo{person}{Yu Zhang}, \bibinfo{person}{Frank~F. Xu}, \bibinfo{person}{Sha Li}, \bibinfo{person}{Yu Meng}, \bibinfo{person}{Xuan Wang}, \bibinfo{person}{Qi Li}, {and} \bibinfo{person}{Jiawei Han}.} \bibinfo{year}{2019}\natexlab{}.
\newblock \showarticletitle{HiGitClass: Keyword-Driven Hierarchical Classification of GitHub Repositories}. In \bibinfo{booktitle}{\emph{2019 {IEEE} International Conference on Data Mining, {ICDM} 2019, Beijing, China, November 8-11, 2019}}, \bibfield{editor}{\bibinfo{person}{Jianyong Wang}, \bibinfo{person}{Kyuseok Shim}, {and} \bibinfo{person}{Xindong Wu}} (Eds.). \bibinfo{publisher}{{IEEE}}, \bibinfo{pages}{876--885}.
\newblock
\urldef\tempurl%
\url{https://doi.org/10.1109/ICDM.2019.00098}
\showDOI{\tempurl}


\end{thebibliography}

%
\clearpage
\pagebreak
\setcounter{section}{0}
\renewcommand\thesection{\Alph{section}}
\section*{Appendix}
\section{Walkthrough}
In our repository\footnote{\href{https://github.com/SasCezar/AutoFL}{https://github.com/SasCezar/AutoFL}}, we provide a \README file with the installation and more detailed info about \autofl. Here we present a short introduction to the setup and use of the tool.
A walkthrough video is also available\footnote{\href{https://youtu.be/ZYWZdYcip2A}{https://youtu.be/ZYWZdYcip2A}}.

\subsection{Installation}
After cloning the repository, we can run the tool. The tool uses \textit{docker} and \textit{docker compose} (version 25.0.0). Therefore, minimal setup is needed. Here are the commands to run:
\begin{lstlisting}[language=bash,caption={Installation},basicstyle=\footnotesize]
git clone --recursive git@github.com:SasCezar/AutoFL.git AutoFL
docker compose up
\end{lstlisting}
The process requires around 10 minutes (depending on the connection) to setup and install all the dependencies.

\subsection{Usage}
\begin{figure*}[htbp!]
    \centering
    \includegraphics[width=.9\textwidth]{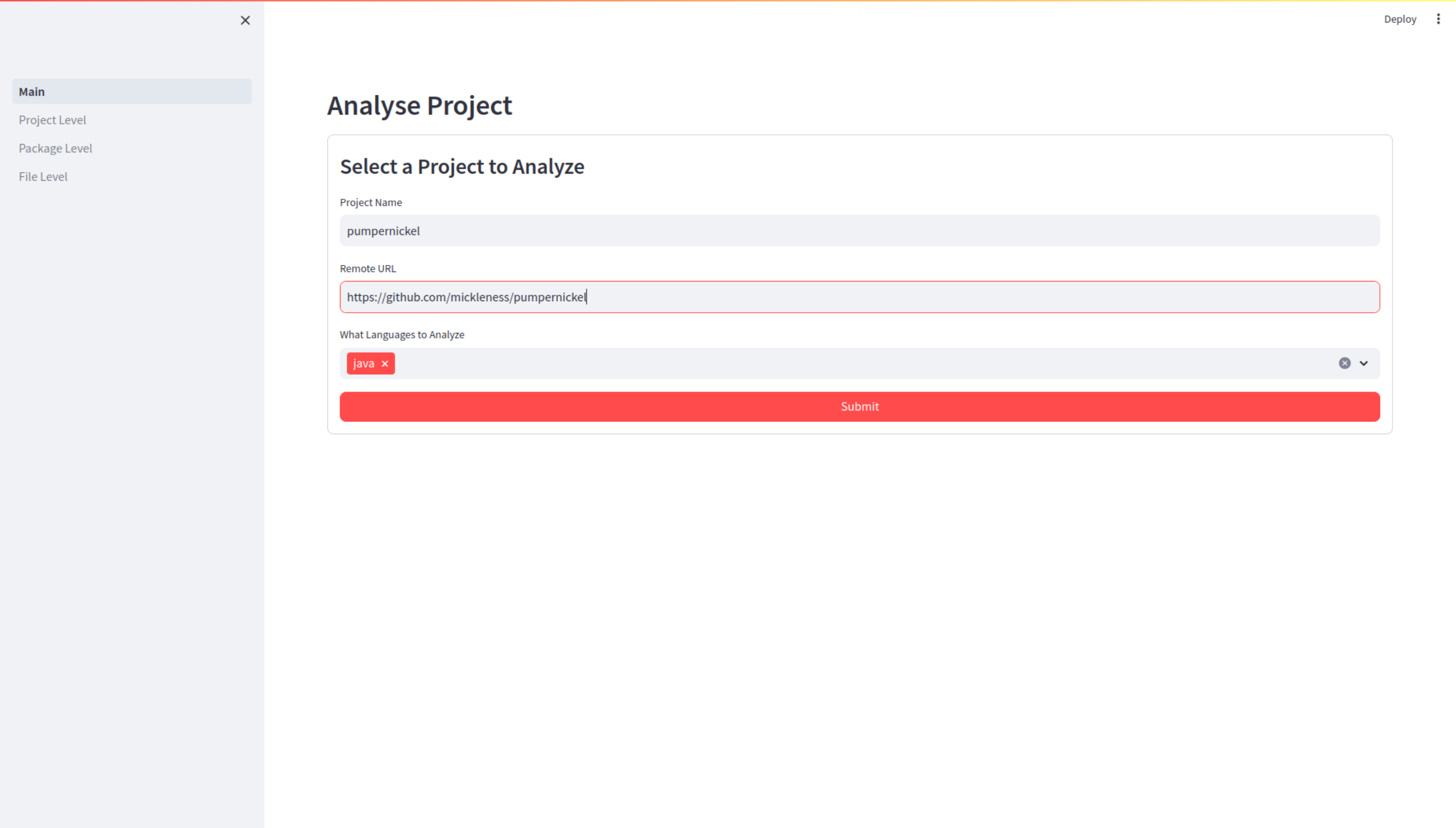}
    \caption{\autofl homepage UI.}
    \label{fig:home_ui}
\end{figure*}

Once the docker containers are up and running, we can access the web UI at the \href{http://localhost:8501/}{http://localhost:8501/} address when running locally. 
The homepage (Figure~\ref{fig:home_ui}) allows for the selection of the project that we want to analyze, including the name, remote URL, and programming language.

\begin{figure*}[htbp!]
    \centering
    \includegraphics[width=.9\textwidth]{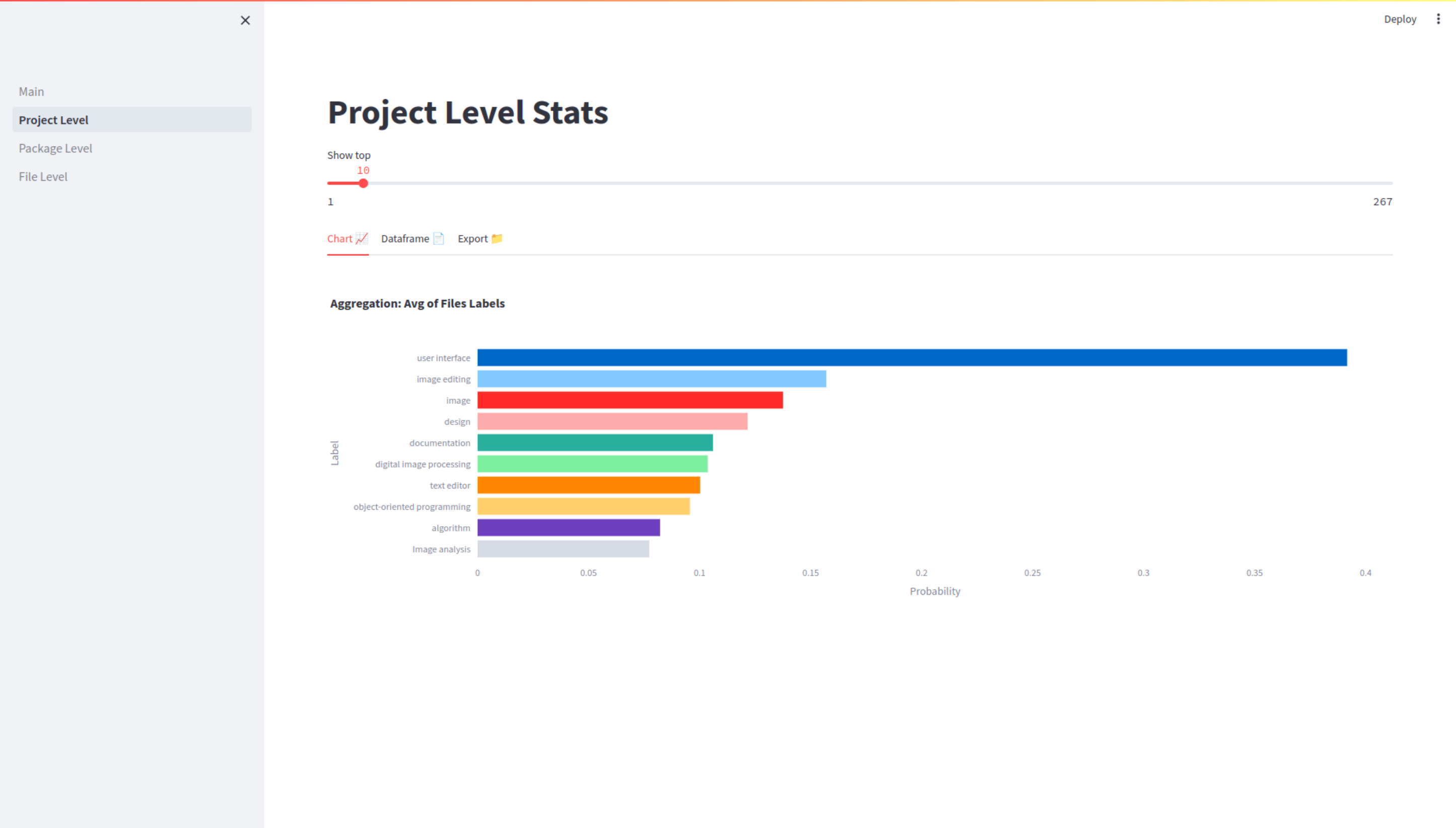}
    \caption{\autofl Project-level results UI.}
    \label{fig:project_ui}
\end{figure*}

\begin{figure*}[htbp!]
    \centering
    \includegraphics[width=.9\textwidth]{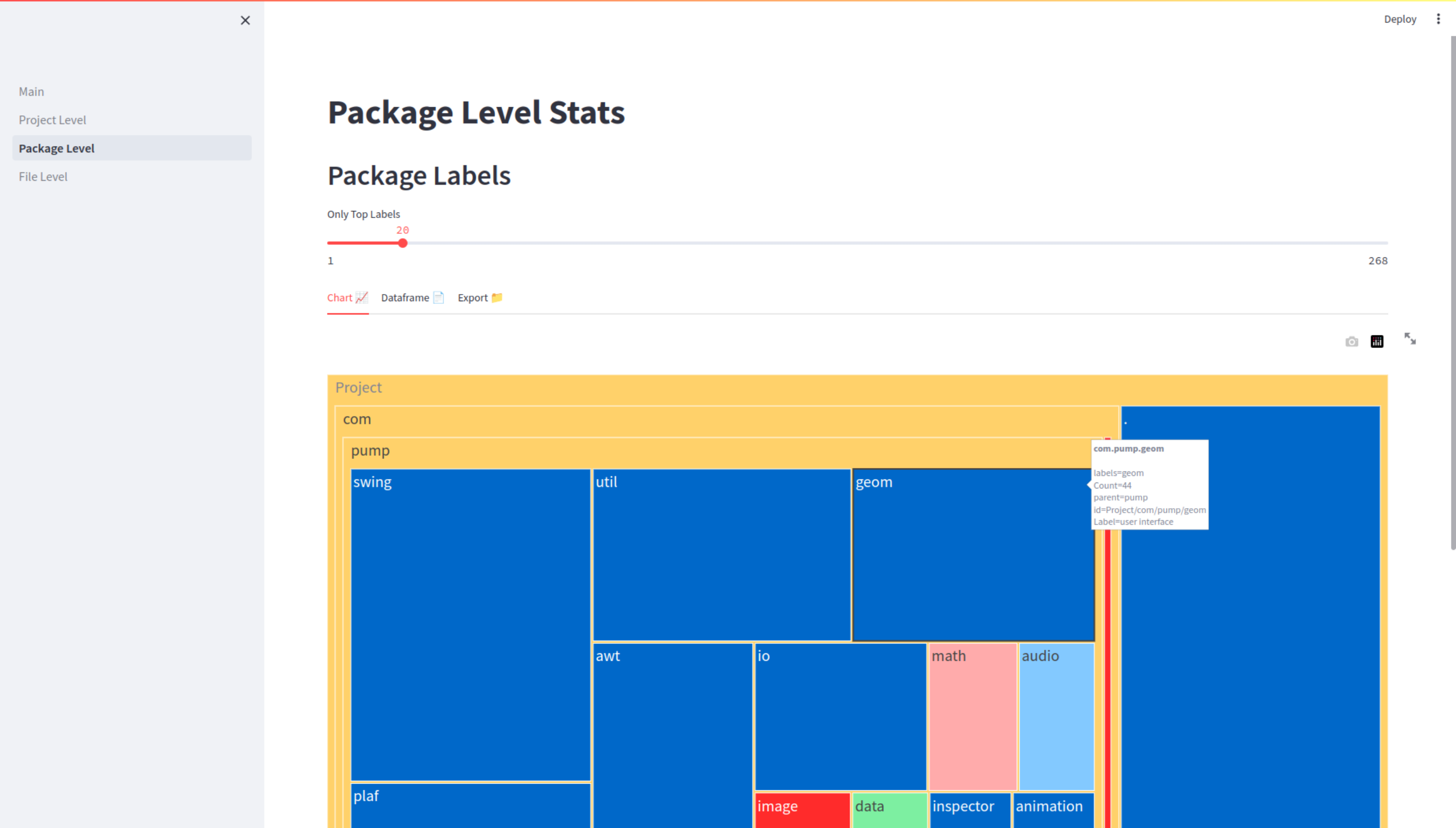}
    \caption{\autofl Package-level results UI.}
    \label{fig:package_ui}
\end{figure*}

\begin{figure*}[htbp!]
    \centering
    \includegraphics[width=.9\textwidth]{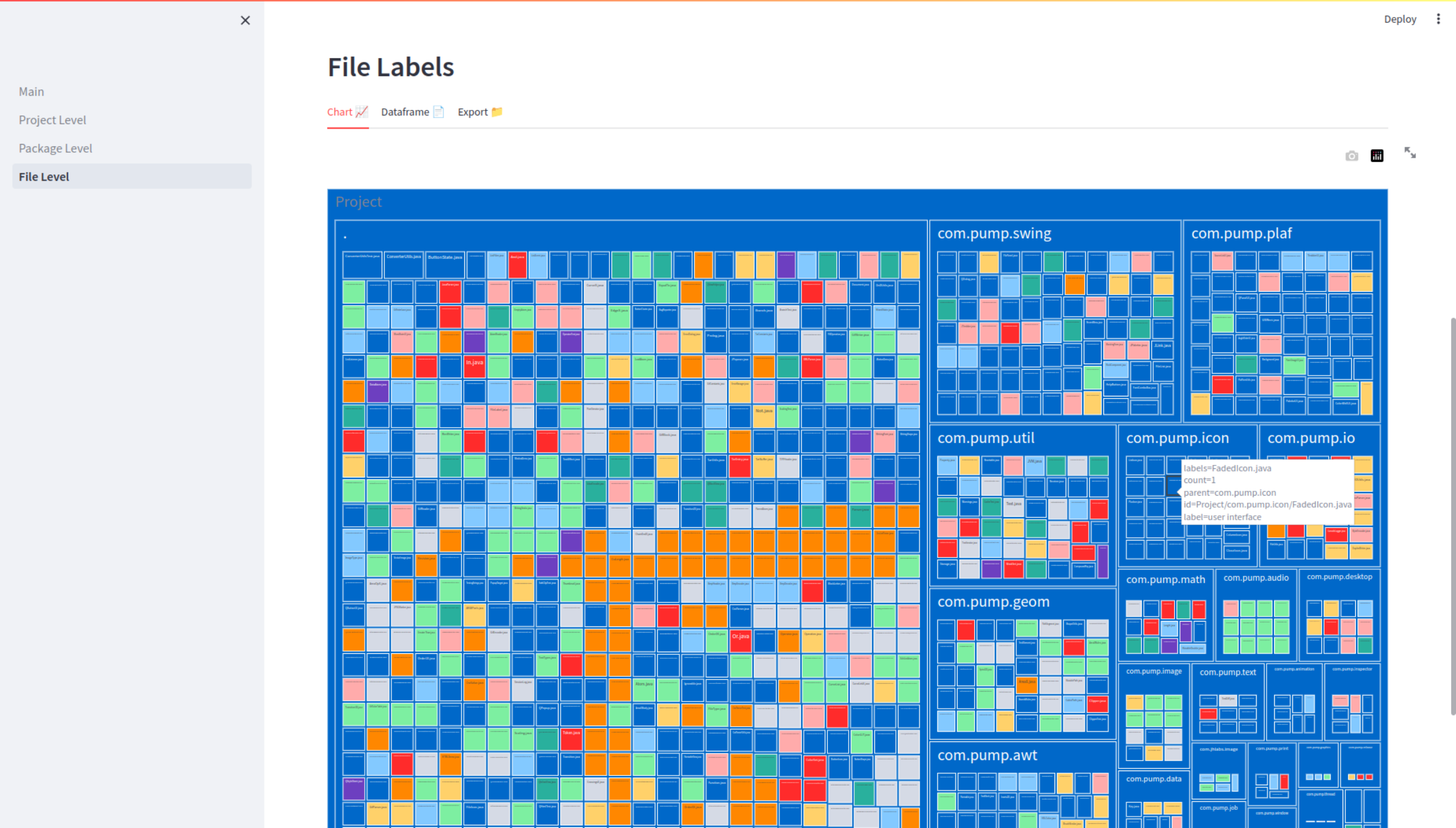}
    \caption{\autofl File-level results UI.}
    \label{fig:file_ui}
\end{figure*}

Once the analysis is completed, the results are available through the side menu. The first menu item brings us to the project-level results (Figure~\ref{fig:project_ui}), where we can see the probabilities for the best labels from the analyzed project. On the second page, the package-level results are displayed (Figure~\ref{fig:package_ui}). We can see the structure and labels for the identified project packages. Finally, in the last menu, we can see the file-level results (Figure~\ref{fig:file_ui}), where the files are the smallest squares, and they are grouped by package.
\pagebreak

\end{document}